\documentclass[a4paper]{article}
\usepackage{graphicx}
\usepackage{latexsym}
\usepackage{amssymb}
\usepackage{amsmath}
\usepackage{multicol}
\usepackage{caption}
\usepackage{subcaption}
\usepackage{float}
\usepackage[round]{natbib}
\usepackage{color}
\definecolor{armygreen}{rgb}{0.19, 0.53, 0.43}
\definecolor{atomictangerine}{rgb}{1.0, 0.6, 0.4}

\begin{document}

\newtheorem{theorem}{Theorem}[section]
\newtheorem{corollary}{Corollary}[theorem]
\newtheorem{lemma}[theorem]{Lemma}

\title{A Network Model for Multiple Selection Questions in Opinion Surveys}
\author{Benati, Puerto}
\author{Stefano Benati\\
Dipartimento di Sociologia e Ricerca Sociale\\
School of International Studies \\
Universit\`{a} di Trento\\
Via Verdi 26, 38122 Trento, Italy
\and Justo Puerto \\
IMUS. Universidad de Sevilla \\
Avda. Reina Mercedes s/n, 41012 Sevilla, Spain}

\date{\today}

\maketitle

\begin{abstract}
Opinion surveys can contain closed questions to which respondents can give multiple answers. We propose to model these data as networks in which vertices are eligible items and arcs are respondents. This representation opens up the possibility of using complex networks methodologies to retrieve information and most prominently, the possibility of using clustering/community detection techniques to reduce data complexity. We will take advantage of the implicit null hypothesis of the modularity function, namely, that items are chosen without any preferential pairing, to show how the hypothesis can be tested through the usual calculation of $p$-values. 
We illustrate the methodology applying it to Eurobarometer data. There, a question about national concerns can receive up to two selections. We will show that community clustering groups together concerns that can be interpreted in consistent way and in general terms, such as Economy, Security and Welfare issues. Moreover, we will show that in this way cleavages between social sectors can be determined.
\end{abstract}

\bigskip

\pagebreak

\section{Introduction and Model Motivation}

Some surveys contain closed questions in which respondents are proposed with a list of items among which they can elicit more than one answer. For example, see \cite{ZA6928}, the Eurobarometer standard survey formulates a question about citizens' concerns in this way:\bigskip

{\it What do you think are the two most important issues facing your country at the moment?} (Max. 2 answers)

\begin{multicols}{2}
\begin{itemize}
\item Crime
\item Rising prices, inflation
\item Taxation
\item Unemployment
\item Terrorism
\item Housing
\item Government debt
\item Immigration
\item Health and social security 
\item The education system
\item Pensions
\item The environment, climate and energy issues
\item Economic situation
\item Other 
\item None 
\item Don't know
\end{itemize}
\end{multicols}

There is a subtle but hidden problem when this kind of questions are to be analyzed. Answers are usually coded and then reported as the frequency by which {\it one} single item has been mentioned or not, see \cite{Rouet2016, Bevan2016, Traber2022} or the a Eurobarometer report such as \cite{Brussels2018}, but actually respondents gave {\it two} answers. If pairs are broken, then reports could loose analytical detail, as a respondent answering the pair \textit{Immigration, Crime} could be substantially different from a respondent answering \textit{Immigration, Unemployment}: in the former case, the concern about immigration is rooted on its effect on personal security, while in the latter case it is rooted on the economic downturn. 
One may argue that the best way to analyze these data is to keep the answers in pair, but it can be problematic. There is a combinatorial explosion of all the possible answers an individual can give as they are all possible pairs from the set of 13 items and one must find some other convenient way of data reduction to continue with the statistical analysis. 

In this contribution, we will propose a network model to represent survey data coming from multiple selection survey questions. We will call this network the Items Graph as it is composed by nodes representing question choices/items and (multiple) arcs between nodes representing the actual answers by respondents. The graph contains as many arcs as the survey respondents, with possible loops corresponding to answers in which just one item has been elicited. Representing data as a network allows next the possibility of using all the tools developed for complex network analysis, such as the use of centrality measures, see \cite{Das2018}, the core-periphery segmentation, see \cite{Tang2019}, or the community detection, see \cite{Fortunato2016}.

Taking inspiration from \cite{Bevan2016}, we will show how to apply community detection to the items graph. In that paper, authors empirically aggregated citizens' concerns in few classes, that are used to detect whether there is a correspondence between issues that are considered important at the national, personal, and European level. Concerns were aggregated following a simple rule-of-thumb, without using any quantitative analysis. Indeed, the use of community detection models can be useful as in the items graph communities are subset of concerns that respondents deemed as homogeneous. In this way a qualitative way of aggregating data is replaced by a quantitative one. 

A peculiar advantage of community detection models is that the detected clusters can be validated by statistical inference. Indeed, a feature of modularity clustering is that it uses an implicit null hypothesis, that assumes that there is no preferential pairings between nodes/items. Its formal definition is delayed to the following section, however, preferential pairings appears when some nodes/items pairs are mentioned more often than what is expected by independent probabilities. Applying the methodology proposed in \cite{Zhang2017}, we will show how to use this null hypothesis for statistical testing and to determine if the communities resulting from modularity optimization are significantly different from random communities. The methodology that we will describe is how to calculate test $p$-values from modularity optimization.

Finally, we will use the Eurobarometer question about the most important national issue to make an exercise with the proposed methodology. It will be seen that clusters are composed of concerns that are logically consistent, that they are statistically validated, that is, they are not a mere effect of chance, that they can be used to characterized how different population segments are characterized by different concerns.

The paper is organized as follows. Section 2 is devoted to introduce the items graph and to prove some of its properties. Section 3 presents the clique partition model for modularity maximization, there it is also presented how to develop an inferential model based on click partition. Section 4 is devoted to our application of the newly developed methodology to the Eurobarometer data. Finally, Section 5 draws our conclusions and future research lines on the topic. 

\section{The Items Graph}

The network model to represent a multi response question is defined as follows. Let $l_1, \ldots, l_n$ be the labels assigned to the answers of the multiple response question. The Items Graph $G = (V, E)$ is composed of the node set $V = \{1, \ldots, n\}$, corresponding to labels let $l_1, \ldots, l_n$, and there is an arc $(i, j)\in E$ for every respondent that answered the $l_i, l_j$ pair. If the answer is a single item $l_i$, then the arc is a loop $(i,i) \in E$. The degree $\delta_i$ of a node $i$ is the number of arcs incident to $i$, with loops counted twice. Note that $G$ contains multiple arcs and multiple loops. Moreover, let $m_{ij}$ be the number of the $i,j$-pair answers, and let $m_i$ be the number of $i$-single answers, then $\delta_i = 2\,m_i + \sum_{j:j \neq i} m_{ij}$.  Let $m$ be the number of respondents, then $|E|=m$.

It can be conjectured that the graph $G$ has a structure that can be revealed by clustering. That is, items $l_1, \ldots, l_n$ could be interpreted as specific expressions of latent variables, expressing preoccupation about some main and general issue, for example the Economy, the Security, the Social Welfare,and so on. Therefore, from the operational point of view, items $l_1, \ldots, l_n$ can be clustered into homogenous groups using an appropriate clustering model. Here, we propose a clique partitioning model with a modularity objective function.

\subsection{Modularity as independence in multiple response questions}

Define $X_i\cup X_j$ as the event that a respondent elicited the $l_i,l_j$ pair, with the notation $X_i\cup X_i$ denoting the event that the respondent elicited $l_i$ as a single answer, then $\Pr[X_i\cup X_j]$ is the probability the one respondent elicited the $l_i,l_j$ pair. Interpreted in the items graph, $\Pr[X_i\cup X_j]$, with possibly $i = j$, is the probability that an arc $(i,j)\in E$. Define $\Pr[X_i]$ as the probability that item $l_i$ is one of the elicited item by a respondent and define $\Pr[X_i|X_j]$ as the conditional probability that a respondent has chosen $l_i$, giving that he or she elicited $l_j$. If $\Pr[X_i|X_j] \neq \Pr[X_i]$, then we will say that there is a {\it preferential pairing} between items $l_i$ and $l_j$. Depending on the difference between the probability, we can asses that the pair $l_i, l_j$ has been elicited more or less frequently than expected. 

Now we consider the following problem: What is the expected number $r_{ij}$ of respondents that selected the $l_i,l_j$ pair under the hypothesis that there is no preferential pairing? Or, interpreted in the items graph, what is the expected number of arcs between the pair $i,j$ under the condition of independence? 

Consider the survey graph $G=(V,E)$ and an
auxiliary oriented graph $G' = (V, E')$, in which
for every non-oriented  edge  $ij \in E$ there are two oriented arcs $(i,j)$ and $(j,i)$ in $E'$ and, if the arc is a loop, for every non-oriented loop $(i,i)$ there are two oriented loops, say $(i,i)^+$ and $(i,i)^-$. So, $|E'| = 2m$. Between $G$ and $G'$ there is the following connection: Let $\cal{A}$ be the event of selecting one edge at random from $E$, let $\cal{E}$ be the event of selecting one arc at random from $E'$. We have:

\begin{equation}\label{c1}
\Pr[{\cal A}= ij ] = \Pr [ {\cal E} = (j,i)] + \Pr[{\cal E} = (i,j)] = 2\Pr[{\cal E} = (i,j)]
\end{equation}

Observe that the number of arcs of $E'$ leaving a node $i$ is $\delta_i$, exactly as the number of entering arcs. If we draw at random on arc $e$ from $E'$, then we have (calculated as the ratio between favorable and possible cases):

\begin{align}
\Pr[ \mbox{$e$ leaves $i$}] = \frac{\delta_i}{2m}\\
\Pr[ \mbox{$e$ enters $i$}] = \frac{\delta_i}{2m}
\end{align}

If there is no preferential pairing, e.g. independence, then:

\begin{align*}
\Pr[{\cal E} = (i,j)]& = \Pr[ \mbox{the arc leaves $i$}]\Pr[\mbox{the arc enter $j$}|\mbox{the arc leaves $i$}]]\\
& = \Pr[ \mbox{the arc leaves $i$}]\Pr[ \mbox{the arc enter $j$}]\\
& = \frac{\delta_i}{2m}\frac{\delta_j}{2m} = \frac{\delta_i \, \delta_j}{4m^2}
\end{align*}

Then we can state:

\begin{theorem}: Under the hypothesis of no preferential matching, the number $r_{ij}$ of expected respondents of pair $l_i,l_j$ is $r_{ij} = \frac{\delta_i \delta_j}{2 m}$.
\end{theorem}

{\bf Proof:} From equation (\ref{c1}), $\Pr[{\cal A}= ij] = \frac{\delta_i \delta_j}{2 m^2}$ and then, as there are $m$ edges in $G$, the expected number of edges is $r_{ij} = \frac{\delta_i \delta_j}{2 m}$.

As a consequence of the Theorem, we can establish the difference
$m_{ij} - r_{ij}$ as the measure of dissimilarity between the actual number of $ij$-respondents with the theoretical one, under the assumption that there is no preferential pairing between items. Actually, this measure is at the core of the modularity index, see \cite{Newman2004}, that is used for community detection in social networks. The only difference is that there is no loops neither multiple arcs in social networks, so that values $m_{ij}$ are an approximation of a null hypothesis, while in our case they are an exact value. Modularity maximization coincides with the clique partition problem, see \cite{Agarwal2008}, and it is revised in the following subsection.

\section{The Clique Partitioning/Modularity maximization model}

The Clique Partitioning (CP) problem is one of the cornerstone of combinatorial optimization, see \cite{Grotschel1989, Grotschel1990}. It can be formulated as follows:
Let $G = (V, E)$ be a complete graph. Let $c_{ij}$ be the similarity measure between node $i$ and node $j$, with $c_{ij}$ being possibly positive, denoting similarity, and negative, denoting dissimilarity. Let $P = \{C_1, C_2, \ldots,C_q\}$ be a feasible partition of $V$ and let $\sum_{(i,j) \in C_k} c_{ij}$  be the sum of the similarity and dissimilarity between vertices of group $k$.  The CP problem consists of finding the node partition $P = \{C_1, C_2, \ldots,C_q\}$ to maximize the objective function 
$\sum_{C_k \in P} \sum_{(i,j) \in C_k} c_{ij}$.  Its Integer Linear Programming (ILP) formulation is: Let $x_{ij}$ be binary variables such that $x_{ij} = 1$ if node $i$ and $j$ are in the same cluster, 0 otherwise.  Then the ILP formulation is: 
	  	
\begin{equation}
z(G) = \max \sum_{i = 1}^{n-1} \sum_{j = i + 1}^n c_{ij} x_{ij}
\label{ilp:fo}
\end{equation}   	
		
\begin{equation}		
x_{ij} + x_{jk} - x_{ik} \le 1, \mbox{ for all $i,j,k \in V$,} i < j < k,  
\label{ilp:c1}
\end{equation}   	
		
\begin{equation}		
x_{ij} - x_{jk} + x_{ik} \le 1, \mbox{ for all $i,j,k \in V$,} i < j < k;  
\label{ilp:c2}
\end{equation}   	

\begin{equation}		
-x_{ij} + x_{jk} + x_{ik} \le 1, \mbox{ for all $i,j,k \in V$,} i < j < k;  
\label{ilp:c3}
\end{equation}   	

\begin{equation}
x_{ij} \in \{0,1\} \mbox{ for all $i,j \in V$,} i < j.
\label{ilp:c4}
\end{equation}
\bigskip
		
The objective function \ref{ilp:c1} selects clusters with high internal similarity and calculate the optimal modularity function $z(G)$ of the graph $G$. The triangle constraints \ref{ilp:c1}, \ref{ilp:c2}, \ref{ilp:c3} represent the property that, if $i$ and $j$ are in the same cluster and so are $j$ and $k$, then $i$ and $k$ must also be in the same cluster. Finally, constraints \ref{ilp:c4} restrict variables to be binary. 

From the problem formulation, note that it is important that similarities $c_{ij}$  take positive and negative values, otherwise there is no incentive to discard negative arcs and the best partition would be $P = \{V\}$.  Moreover, the optimal partition $P = \{C_1, C_2, \ldots,C_q\}$ can contain clusters  with positive and negative internal arcs, as long as the sum of the positive similarities overpasses the sum of the negative dissimilarities. Finally, note that the number of clusters $q$ of the partition is not fixed in advance, but it is a problem outcome.

To determine whether the items graph $G = (V, E)$ has a clustered structure, we compare the actual graph with an hypothetical survey graph $G’ = (V, E’)$, having the property of no preferential pairing.  Let $m_{ij}$ be the number of arcs of $E$ connecting two nodes $i$ and $j$, which corresponds to the number of respondents that actually answered the $l_i, l_j$ pair and let $r_{ij} = \frac{\delta_i \delta_j}{2m}$ be the expected number of arcs if there is no matching preferences, then the difference $c_{ij} = m_{ij} - r_{ij}$ is an indicator of the discrepancy between a structured and an unstructured graph: The highest the difference, the most the actual graph $G$ departs from the a theoretical $G’$ having no pairing patterns. Therefore, we can detect cluster of paired items finding groups of nodes with high internal cohesion. That is, the actual graph has a cluster structure $P = \{C_1, C_2, \ldots,C_q\}$, if the number of arcs connecting the nodes within a group $C_k$ are more than what is expected, that is, if $\sum_{(i,j) \in C_k} c_{ij} > 0$,  and this holds true for every $k = 1,\ldots,q$. Therefore, to determine the optimal node clustering the natural choice is to use the CP problem. That is, finding the node partition $P = \{C_1, C_2, \ldots,C_q\}$ such that the objective function 
$\sum_{C_k \in P} \sum_{(i,j) \in C_k} c_{ij}$ is maximized. 

\subsection{Inference with the CP model}\label{s:hiptest}

The graph $G’$ can be considered as the benchmark for a null hypothesis, see \cite{Zhang2017}. There it can be seen how to compare the actual modularity value $z(G)$ (from the objective function (\ref{ilp:fo})) with the theoretical values of $z(G')$ of a graph $G'$ in which there are no preferential pairings. Of course, $z(G')$ under the null hypothesis is characterized by a probability distribution that must be used to calculate the $p$-value of the test. Unfortunately the analytical distribution of $z(G')$ is unknown, but it can be simulated empirically by making a large number of artificial graph $G'$, characterized by no preferential pairing. Let $G_i, i = 1,\ldots,N$ be a i.i.d. sequence of simulated random graphs, $I\{\omega\}$ the indicator function of event $\omega$, then the test $p$-value is approximated by the formula: 

\begin{equation}\label{pvalue}
\hat{p}\mbox{-value} = \frac{1}{N} \sum_{i=1}^N I\{z(G) \le z(G_i)\}
\end{equation}

It remains to describe how $G_i$'s are simulated, that is, what is the formal definition of the null hypothesis. We are using the configuration model, see \cite{Newman2010}: given a graph $G(V,E)$ with $n$ nodes and degree sequence $\delta = (\delta_1, \ldots , \delta_n)$, the null model for the modularity measure is a random graph model having the same degree sequence but without preferential pairings. It can be simuleted by the following operations. Every edge $e =ij$ of the empirical graph $G = (V,E)$ is cut into two parts, say $l_1$ and $l_2$, with $l_1$ incident to $i$ and $l_2$ incident to $j$, called \textit{stubs}. Next, two different stubs are selected randomly and paired and the process repeated for a large number of iterations. The way in which $G'$ is built implies that the degree sequence $\delta$ remains unvaried, but eventual preferential pairings are broken by the random reassignment of stubs. It is worth noting that the typical flaws of the procedure when applied to community detection, namely, theappearance of loops and multiple edges, see \cite{Cafieri2010}, does not apply to items graph, as in the latter multiple edges and loops are allowed.

To summarize, the $p$-value of the test is calculated through the following procedure:

\begin{itemize}
\item Step 1: Calculate $z(G)$, the value of the best CP of objective function \ref{ilp:fo}
\item Step 2: Repeat $i = 1,\ldots,N$ times:
\begin{itemize}
\item Generate a random uniform $G_i$ graph with fixed degree sequence (it can be done with the rewiring method described in \cite{Newman2010}).
\item Calculate $z(G_i)$, the objective function \ref{ilp:fo} of CP applied to $G_i$.
\end{itemize}
\item Use the sample $z(G_i), i = 1,\ldots,N$,  to determine the empiric distribution of $z(G)$ under the null hypothesis.
\item Calculate $p$-value using formula (\ref{pvalue}).
\end{itemize}

In the next experiments, we have used the software GuRoBi, \cite{gurobi2022}, to calculate $z(G)$ by solving the CP problem. Even though the CP problem is NP-complete, the instance size is small as the number of items is 16 at most. So, computational times are negligible even though they must repeated $N$ times (less than 30 seconds when N = 1000, as in the following experiments).

\section{An application: What is salient for public opinions?}

The salience of a political issue is important to political analysis, as it affect both voters' behavior and governments' priorities. Moreover, salience could depend by different cleavages, such as social classes, political position and so on. In the next application it can be seen how the methodology described so far can be applied to discover:

\begin{itemize}
\item whether citizens' concerns can be clustered into homogeneous classes, grouping together concerns having the same latent source;
\item whether citizens worried for the same problem class can be characterized by any social feature, such as age, job position, social class or political position.       
\end{itemize}

For illustrative purposes, we use the Eurobarometer ZA6928 surveyed in November 2017, available in Gesis database \cite{ZA6928}, and we compare three national audiences from Italy, Spain and West Germany. We consider the two selections question: {\it What do you think are the two most important issues facing your country at the moment?}. Here respondents can elicit up to two items among the list reported in the introduction, with the possibility of selecting nothing or just one item. Formally, the answer to the question is a variable $X$ whose outcomes are every pair of issues, for example $x_i =$ {\it Immigration,Crime}, or the outcome is a singleton, such as $x_i =$ {\it Health}. The $X$ domain is composed of as many answers as the item pairs, that is, the 16 items can be combined in more than 90 ways and they are the faithfull representation of the survey data. However, this variable $X$ has never been analyzed in its full complexity. Rather, the data frame containing the survey responses $X$ proceeds with a simplification. It splits $X$ into 16 variables/columns, say $Y_i, i = 1, \ldots, 16$, reporting the dichotomy {\it mentioned/not mentioned} for every single issue. After that, the statistical analysis is usually carried out using the simplified variables $Y_i$, see for example \cite{Rouet2016, Bevan2016, Traber2022}. and one of same Eurobarometer reports such as \cite{Brussels2018}. 

One may argue that the variables $Y_i$ are not the faithful translation of the original question. Giving the possibility of two answers, {\it What are the most important issues} is translated into {\it Is [issue name] among the two most important issues?}. One may claim that passing from $X$ to $Y_i, i = 1,\ldots,16$ results with information lost and it could flaw the following statistic analysis. For example, in the original survey, the {\it Immigration} issue can be mainly combined with {\it Unemployment} or, alternatively, with {\it Crime}. The two possibility leads to a completely different social interpretation of the choice: in the former case, immigration is an issue because it can worsen the job market, while in the latter case it is an issue because it can worsen the public security. Next, suppose that two basic exploratory techniques such as two-way tables and correlations are used to analyze $X$. One cannot apply two-way tables directly to $X$, as it would imply a table with at least 90 lines, so one could use correlation on the simplified data $Y_i$ to reveal whether some issues have been consistently mentioned in pairs. Unfortunately, this is not a viable methodology: due to the constraint on choosing at most two item, a mention (standing for 1 value) is always most often combined with a no mention, (standing for 0 value) and then the correlation is a negative number for all pairs $Y_i, Y_j$.

\begin{figure}[H]\label{f1}
     \centering
		\begin{subfigure}[b]{0.95\textwidth}  
		\centering
		\includegraphics[width=0.95\textwidth]{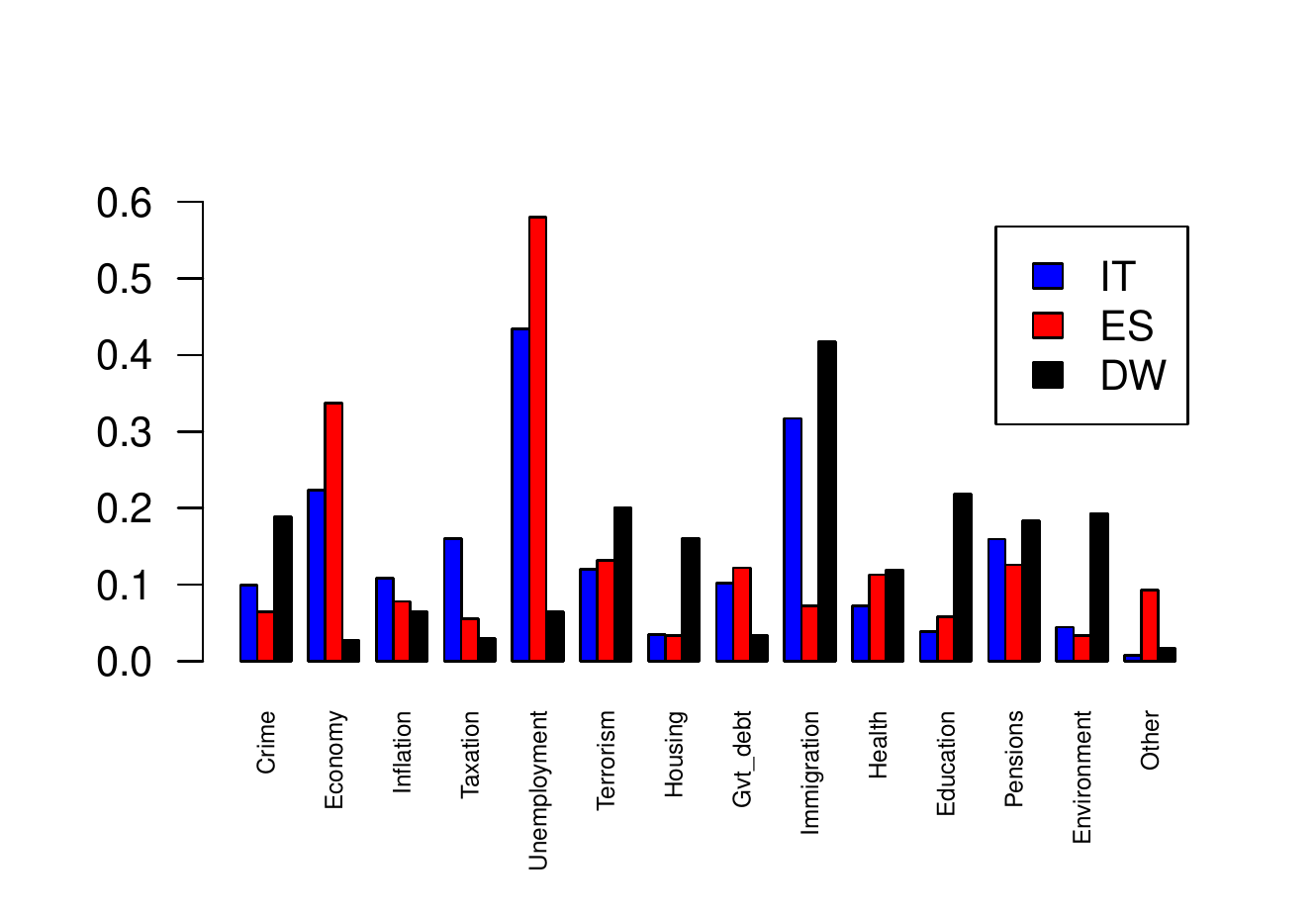}
		\caption{National histogram.}
		\label{f1:h}
		\end{subfigure}
		
     \begin{subfigure}[b]{0.3\textwidth}
         \centering
         \includegraphics[width=\textwidth]{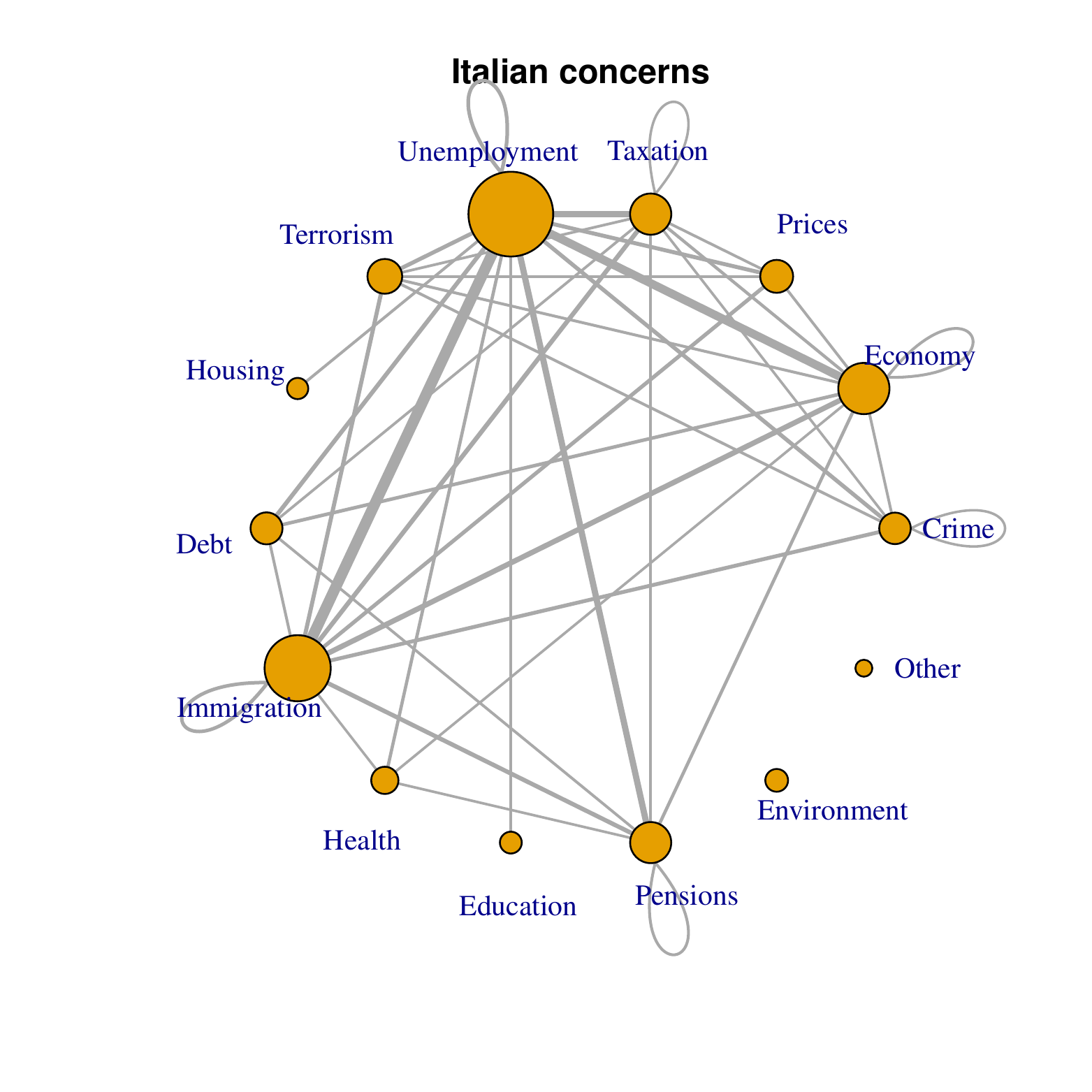}
         \caption{Italian issues.}
         \label{f1:g1}
     \end{subfigure}
     \hfill
     \begin{subfigure}[b]{0.3\textwidth}
         \centering
         \includegraphics[width=\textwidth]{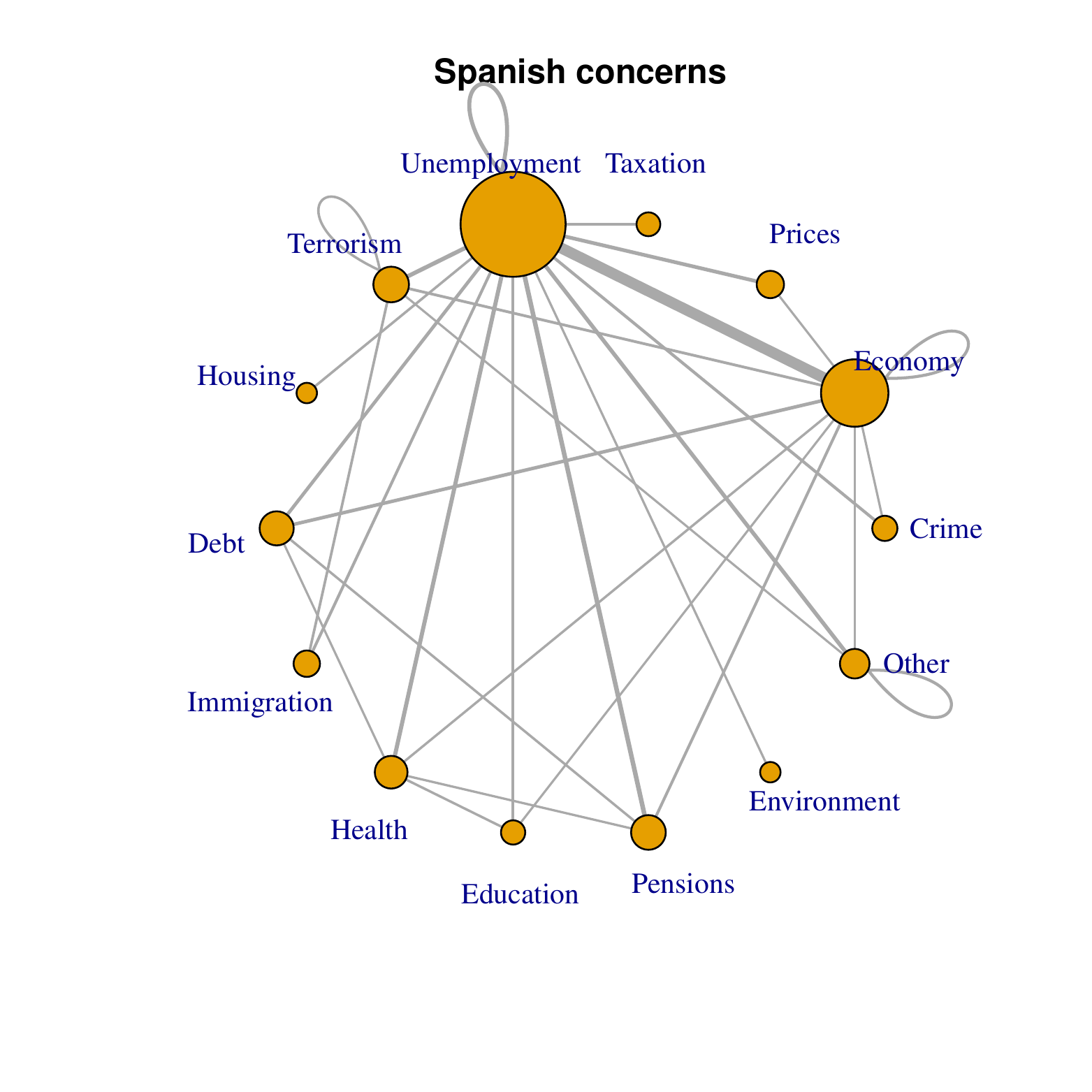}
         \caption{Spanish issues.}
         \label{f1:g2}
     \end{subfigure}
     \hfill
     \begin{subfigure}[b]{0.3\textwidth}
         \centering
         \includegraphics[width=\textwidth]{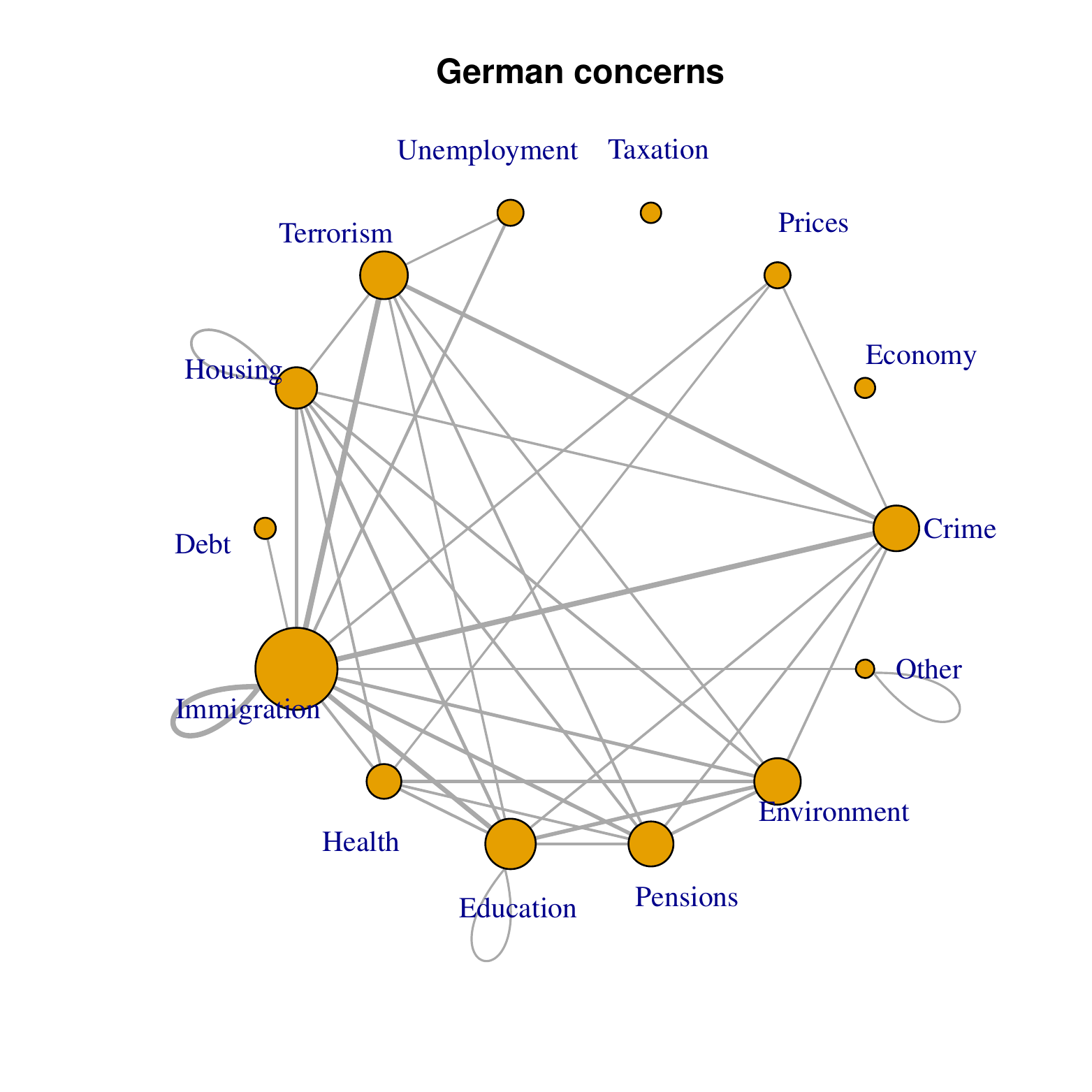}
         \caption{German issues.}
         \label{f1:g3}
     \end{subfigure}
		\caption{Issues histogram and national graphs.}
\end{figure}

We will see how the issues items graph can be used to overcome the difficulty of the aforementioned procedures and then to retrieve information available in $X$. In Figure (\ref{f1:h}) we report the histogram and the  items graph of the three nations. From the histogram, it can be seen that the two most important issues aggregated by countries are Unemployment and Economy for the Spanish, Unemployment and Immigration for the Italians, Immigration and Education for the Germans. Among the less mentioned issues it is remarkable that Pensions is evenly mentioned in the three countries, Environment and Housing are an issue for the Germans only, Taxation for the Italians only, and a not trivial frequency of Spanish reported Other, perhaps referring to the Catalunya dispute that was concurrent to the survey. In the items graphs of Figures (\ref{f1:g1}), (\ref{f1:g2}), (\ref{f1:g3}), data about the item combinations are reported, in the form of the number of respondents that elicited an issues pair. The visualization is provided in such a way that nodes and arcs are larger if more respondents answered that issue or pair of issues, and for graphical purposes arcs smaller than a given threshold are canceled. Regarding the Italian concerns, see Figure (\ref{f1:g1}), it can be seen that most answers lie in the triangle Immigration-Unemployment-Economy, but Pensions seems a well connected issue too. Actually, it is questionable if the weights we are observing are significant, relying on the fact that they results from what we called preferential pairing, or they are just a visual effect resulting from the large number of choices. Note also that some loops are visible: for some respondents there is only one national problem. In Figure (\ref{f1:g2}), The Spanish items graph reveals the connection between Unemployment and Economy, but with Health and Pensions as well, with which Unemployment may form some preferential pairing. Finally, in see Figure (\ref{f1:g3}), the German graph reveals the triangle formed by Immigration-Terrorism-Crime.

\begin{figure}[H]\label{f2}
    \begin{subfigure}[b]{0.3\textwidth}
         \centering
         \includegraphics[width=\textwidth]{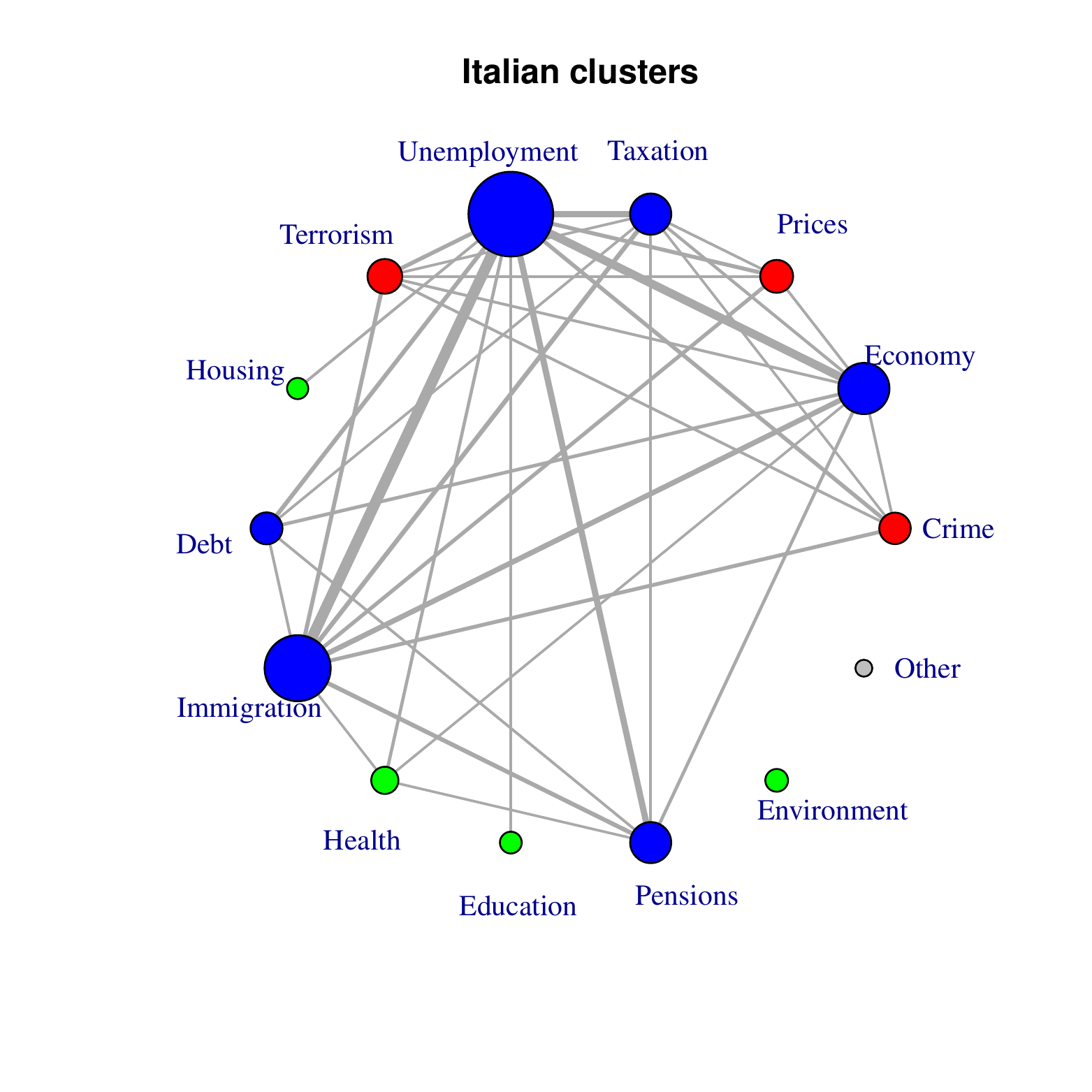}
         \caption{Italian issue clusters.}
         \label{f2:g1}
     \end{subfigure}
     \hfill
     \begin{subfigure}[b]{0.3\textwidth}
         \centering
         \includegraphics[width=\textwidth]{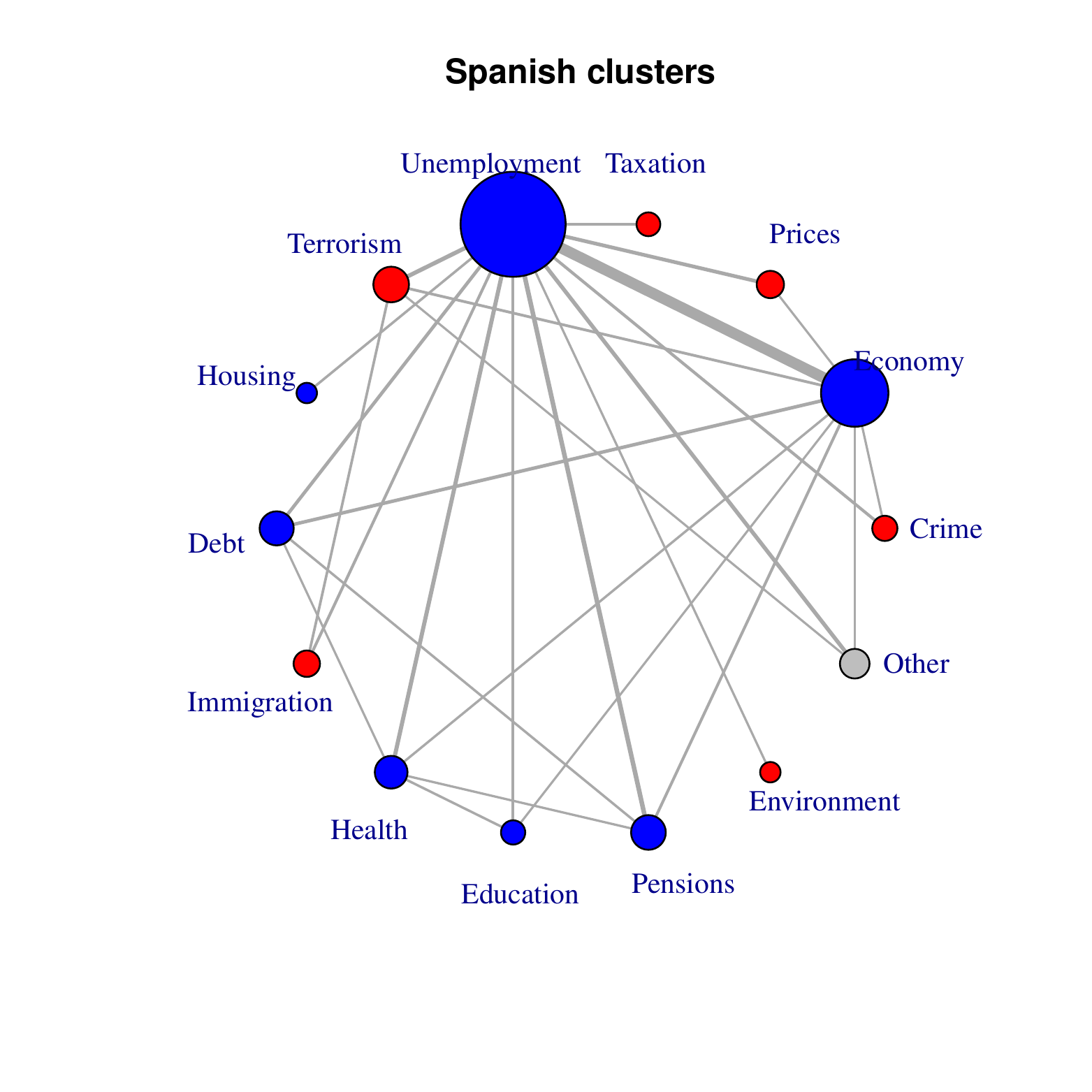}
         \caption{Spanish issue clusters.}
         \label{f2:g2}
     \end{subfigure}
     \hfill
     \begin{subfigure}[b]{0.3\textwidth}
         \centering
         \includegraphics[width=\textwidth]{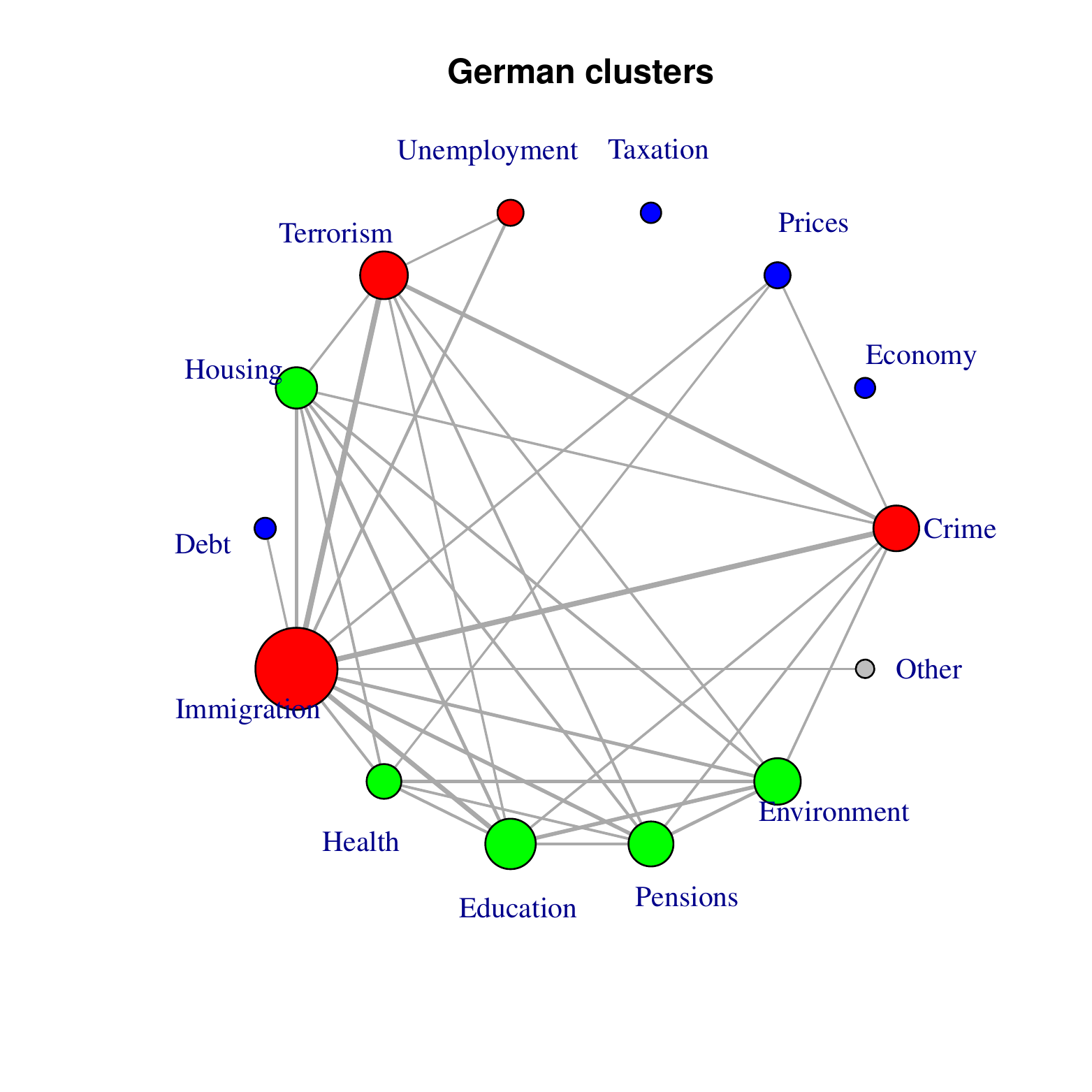}
         \caption{German issue clusters.}
         \label{f2:g3}
     \end{subfigure}
		\begin{subfigure}[b]{0.3\textwidth}
         \centering
         \includegraphics[width=\textwidth]{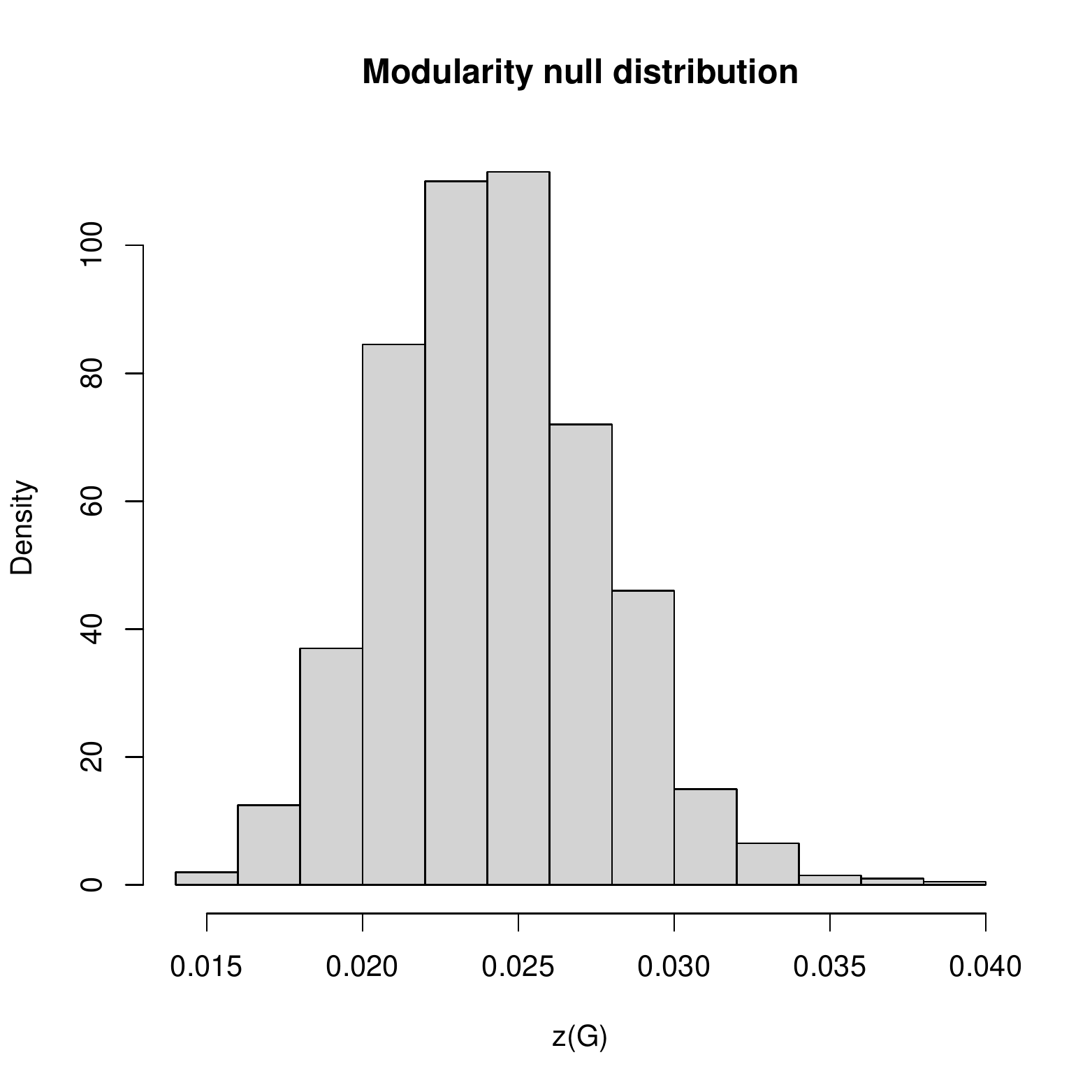}
         \caption{Italian null distribution.}
         \label{f2:h1}
     \end{subfigure}
     \hfill
     \begin{subfigure}[b]{0.3\textwidth}
         \centering
         \includegraphics[width=\textwidth]{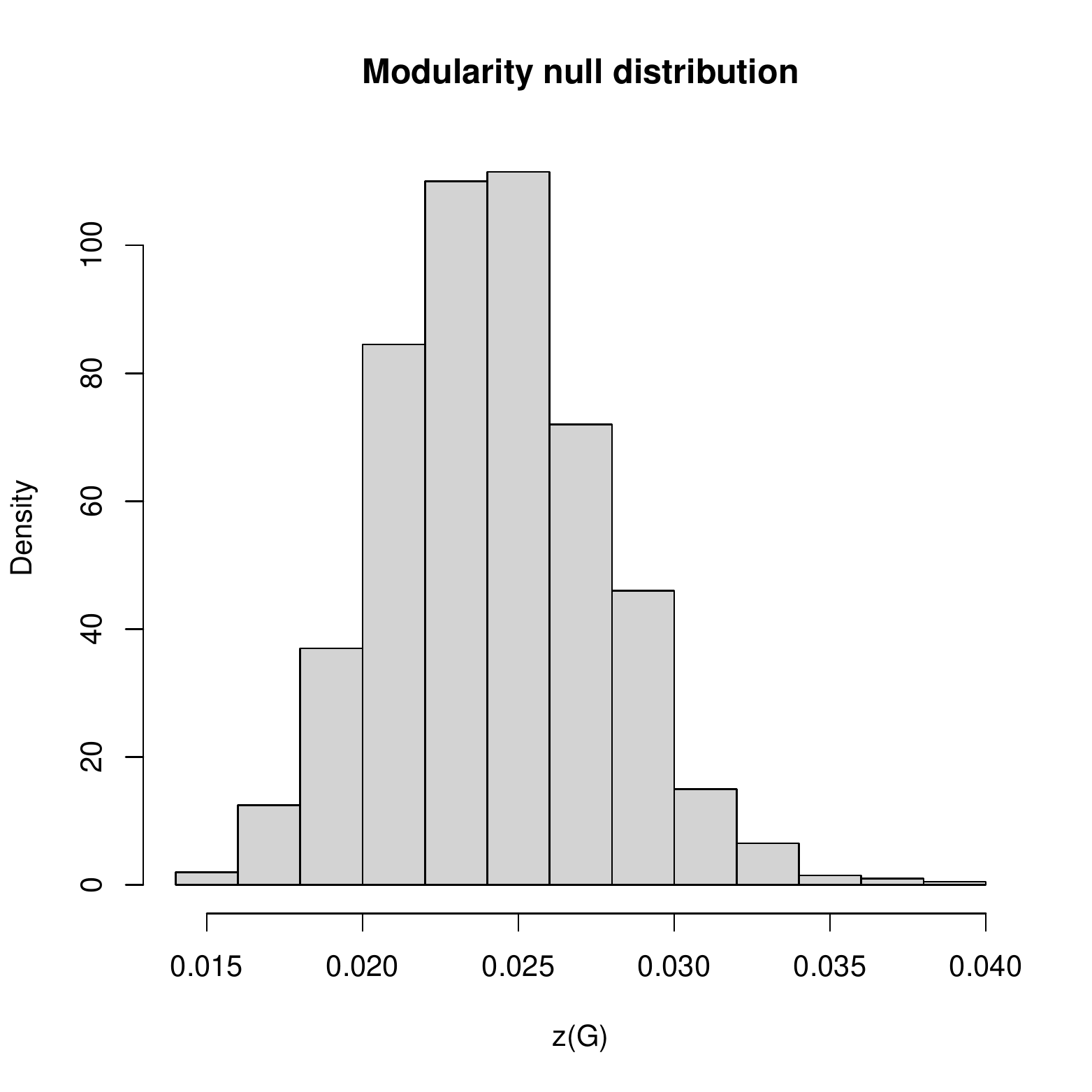}
         \caption{Spanish null distribution.}
         \label{f2:h2}
     \end{subfigure}
     \hfill
     \begin{subfigure}[b]{0.3\textwidth}
         \centering
         \includegraphics[width=\textwidth]{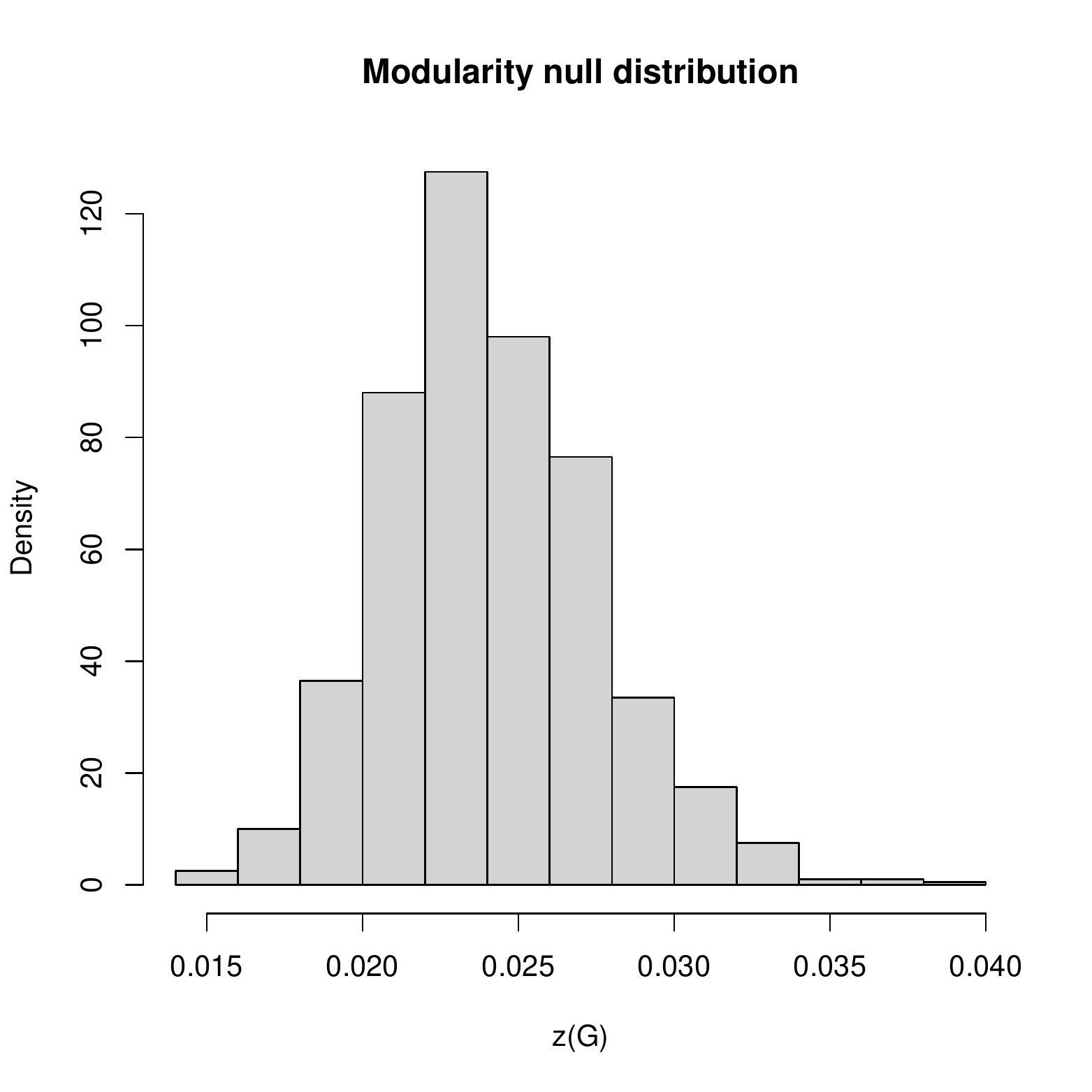}
         \caption{German null distribution.}
         \label{f2:h3}
     \end{subfigure}

		\caption{Issues clusters and relative modularity null distributions.}
\end{figure}

To check whether the most cited issues pair emerged as cases of preferential pairings, we applied modularity optimization. In the case of Italy, we found the optimal modularity value $z(G) = 0.051$ corresponding to the partition reported in Figure (\ref{f2:g1}). Issues are divided into 3 main groups:
\begin{itemize}
\item Group 1: Unemployment, Economy, Immigration, Debt, Taxation, Pensions.
\item Group 2: Crime, Terrorism, Prices.
\item Group 3: Housing, Health, Education, Environment.
\end{itemize}
As can be seen, group 1 is mainly composed by Economic issues, group 2 by Security issues, group 3 by Welfare issues. Note that Immigration is an issue combined with Economic rather than Security issues, and Pensions are combined with the Economy rather than the Welfare.
%Apparently, the partition has a valid interpretation, but it could be questionable whether this partition is the outcome of the preferential pairing between issues, or it is indistinguishable from the null hypothesis, e.g. issues are paired with independent probability. 
As described in Section \ref{s:hiptest}, hypothesis testing is done by network rewiring and simulation, so survey graphs $G_i, i = 1, \ldots, 1000$ are simulated under the null hypothesis of the configuration model, then optimal modularity $z(G^0_i)$ is calculated. Next, we compare the modularity $z(G)$ with the $z(G_i)$ histogram to calculate the experimental $p$-value. For the Italian case, in Figure (\ref{f2:h1}) we reported the histogram of simulated values $z(G_i)$, where it can be seen that the null hypothesis modularity ranges from 0.015 to 0.040. Indeed, given the empiric modularity $z(G) = 0.051$, the experimental $p$-value is 0, rejecting the hypothesis that the items graph resulted from the configuration model. To corroborate this claim, note that the histogram is reminiscent of a bell curve approximately normal and calculating the test $z$-score we obtain 7.78, way larger than any typical hypothesis testing threshold. 
   
The same analysis is repeated for Spain and Germany. As can be seen in Figure \ref{f2:g2}, Spanish concerns are divided in two main groups:
\begin{itemize}
\item Group 1, Security issues: Crime, Prices, Taxation, Terrorism, Immigration, Environment.
\item Group 2, Economic and Welfare issues: Economy, Unemployment, Housing, Debt,      Health, Education, Pensions.
\end{itemize}

Note that Economic issues are merged with the Welfare, while Immigration is included among the Security issues. The empiric modularity is $z(G) = 0.072$, while the $H_0$-modularity ranges from 0.015 to 0.030, see the histogram (\ref{f2:g2}), the test $z$-score is 15.59. Therefore we can reject the null hypothesis of the configuration model.

The German concerns are divided into three main groups:
\begin{itemize}
\item Group 1, Security issues: Crime, Unemployment, Terrorism, Immigration. 
\item Group 2, Economic issues: Economy, Prices, Taxation, Debt.
\item Group 3, Welfare issues: Housing, Health, Education, Pensions, Environment.
\end{itemize}

Note that Immigration and Unemployment are among the Security issues, while the Welfare and the Economic issue are clearly specified by the expected terms. 
The empiric modularity is $z(G) = 0.072$, while the $H_0$-modularity ranges from 0.010 to 0.040, the $z$-score is 16.11. Therefore we can reject the null hypothesis of the configuration model.

In the next analysis, we will show how issue clustering can be applied as a technique of dimensionality reduction, projecting the whole list of issues (including pairs) into two or three classes. Then, we can analyze if respondents whose concerns are within one of the classes can be described in terms of some social cleavage, such as age, job position, and so on. The cleavage that we are going to consider are:

\begin{itemize}
\item Age: We have used the Eurobarometer recoded age into 4 classes: 15-24, 25-39, 40-54, 55 years old and more.
\item Social class: Respondents can locate themselves on a 5-tired social class level that we recoded into Lower and Middle class (Upper class respondents are aggregated as they are never more than a few units).  
\item Job position: We have used the Eurobarometer recoding into Employed, Self-Employed, and Not-Working.
\item Political position: Respondents can locate themselves on a 10-tired political space that we recoded into Left (from 1 to 3), Middle (from 4 to 6), Right (from 7 to 10). 
\end{itemize}  

\begin{figure}[H]\label{f3}
    \begin{subfigure}[b]{0.3\textwidth}
         \centering
         \includegraphics[width=\textwidth]{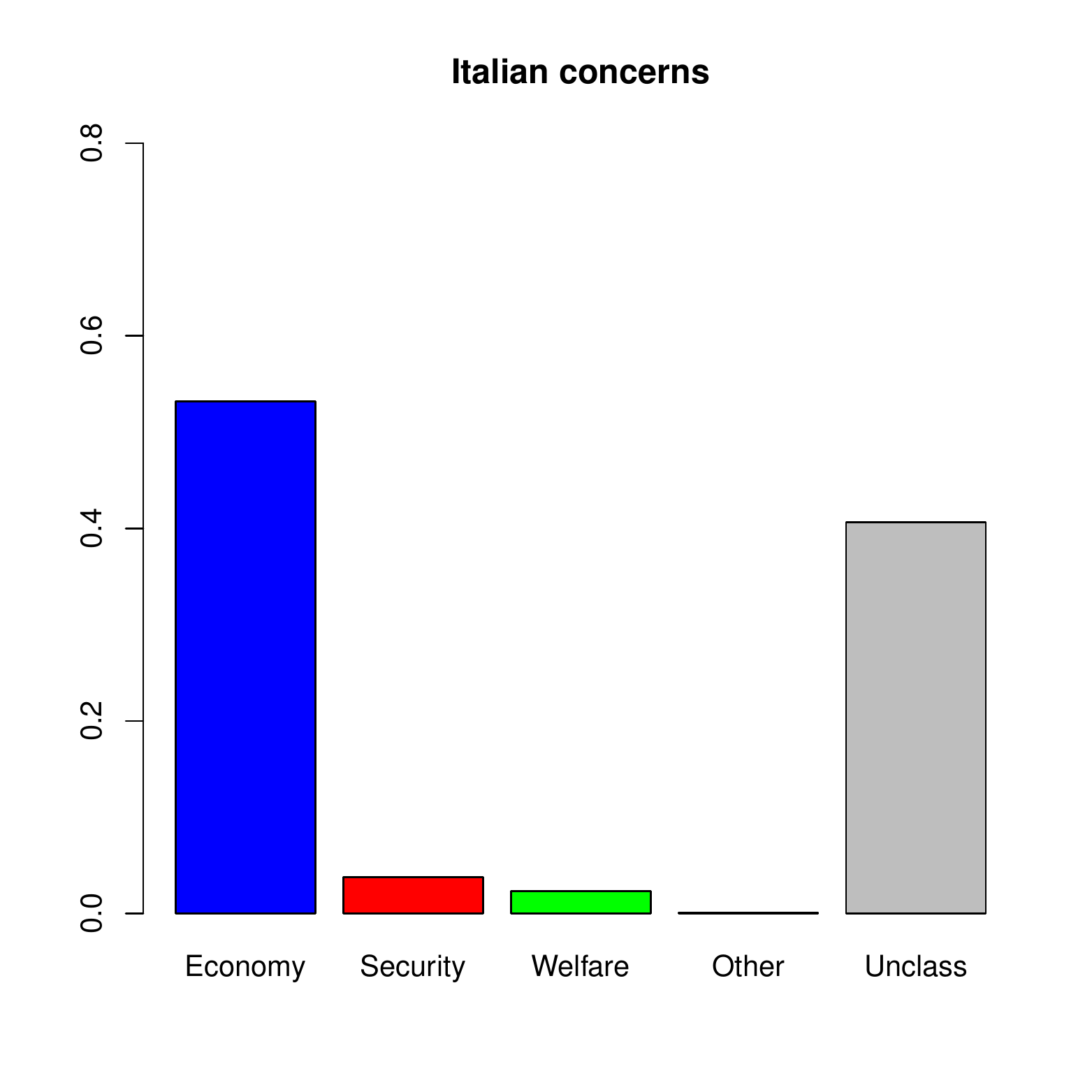}
         \caption{Italian cluster histogram.}
         \label{f3:h1}
     \end{subfigure}
     \hfill
     \begin{subfigure}[b]{0.3\textwidth}
         \centering
         \includegraphics[width=\textwidth]{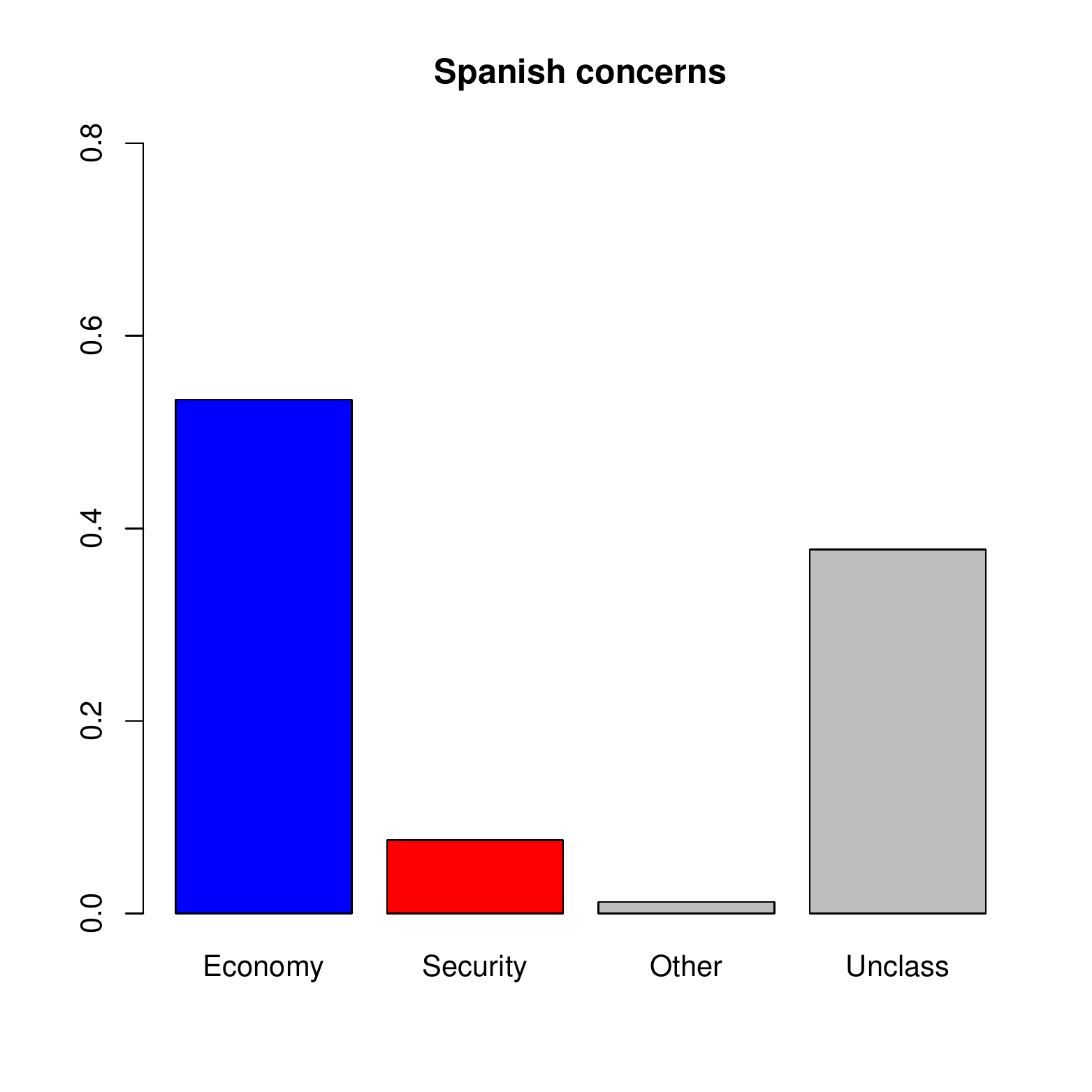}
         \caption{Spanish cluster histogram.}
         \label{f3:h2}
     \end{subfigure}
     \hfill
     \begin{subfigure}[b]{0.3\textwidth}
         \centering
         \includegraphics[width=\textwidth]{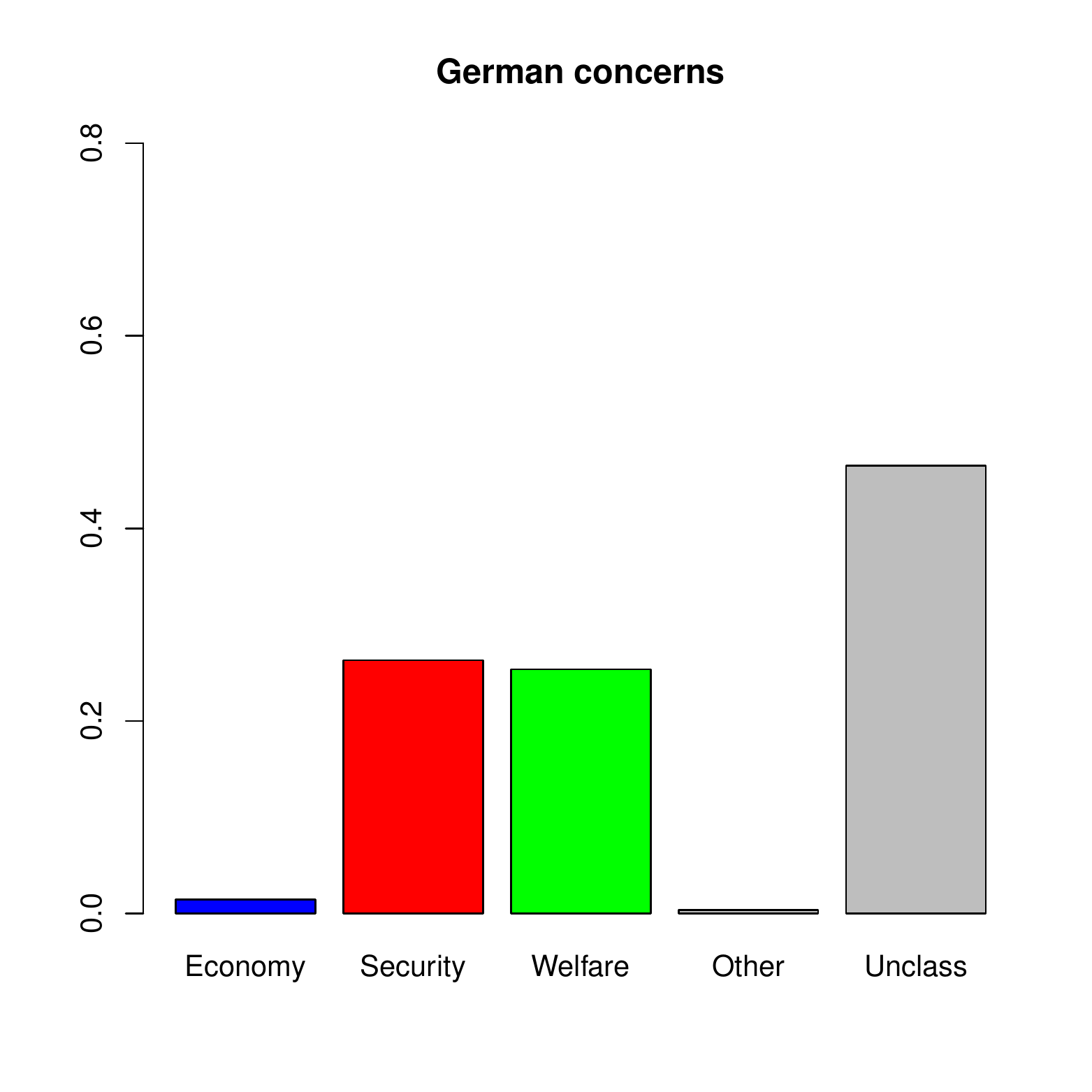}
         \caption{German cluster histogram.}
         \label{f3:h3}
     \end{subfigure}
		\begin{subfigure}[b]{0.3\textwidth}
         \centering
         \includegraphics[width=\textwidth]{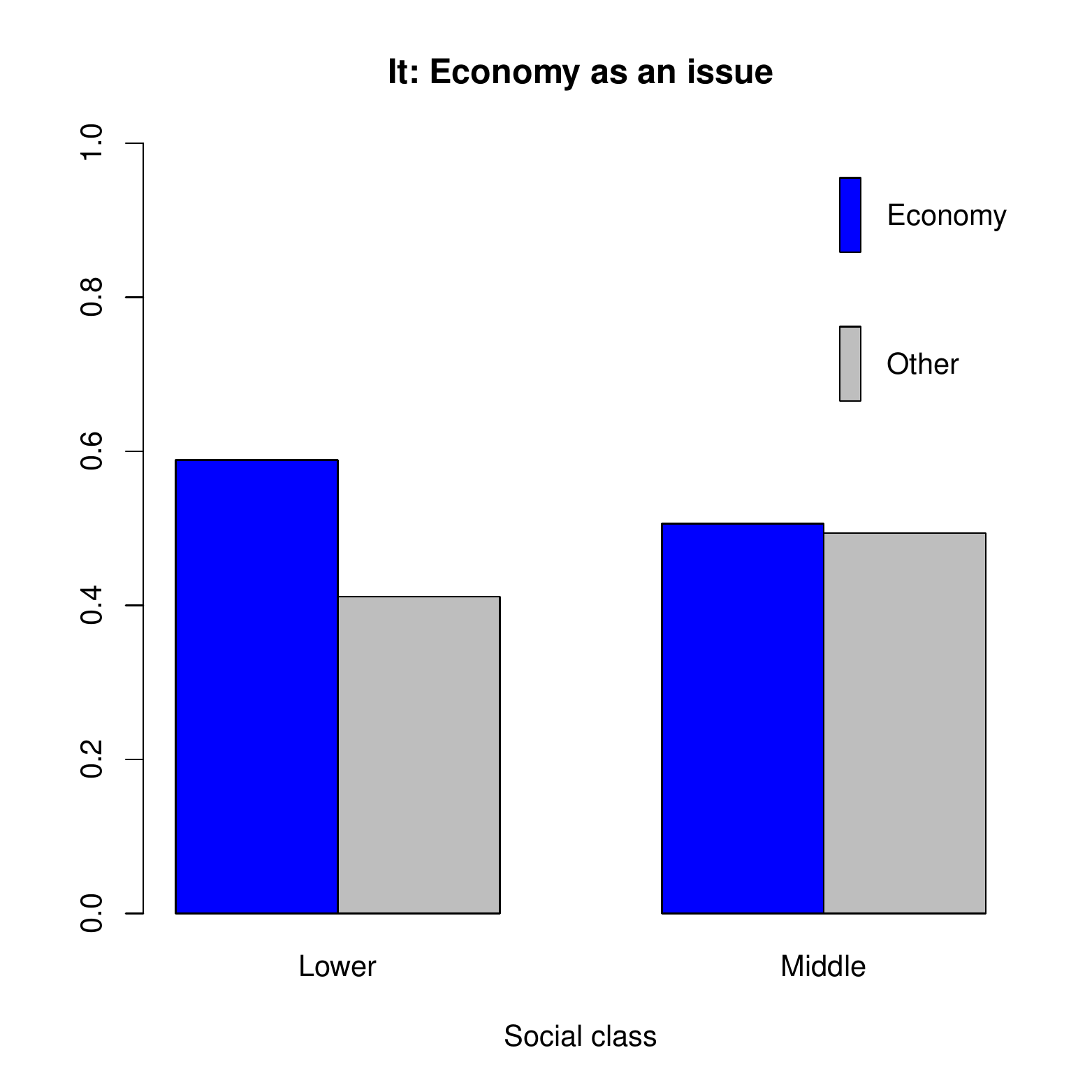}
         \caption{Italian two-ways table.}
         \label{f3:t1}
     \end{subfigure}
     \hfill
     \begin{subfigure}[b]{0.3\textwidth}
         \centering
         \includegraphics[width=\textwidth]{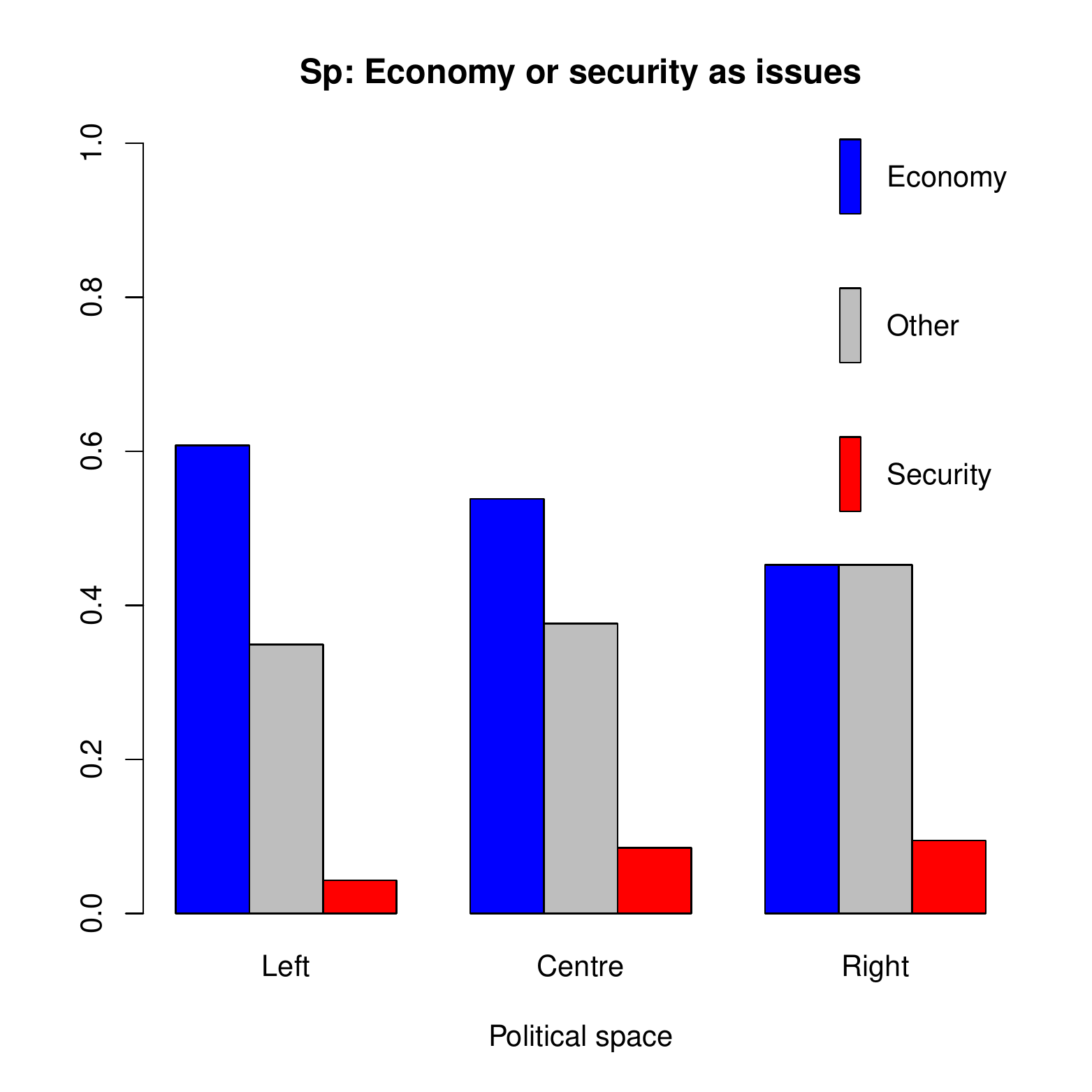}
         \caption{Spanish two-ways table.}
         \label{f3:t2}
     \end{subfigure}
     \hfill
     \begin{subfigure}[b]{0.3\textwidth}
         \centering
         \includegraphics[width=\textwidth]{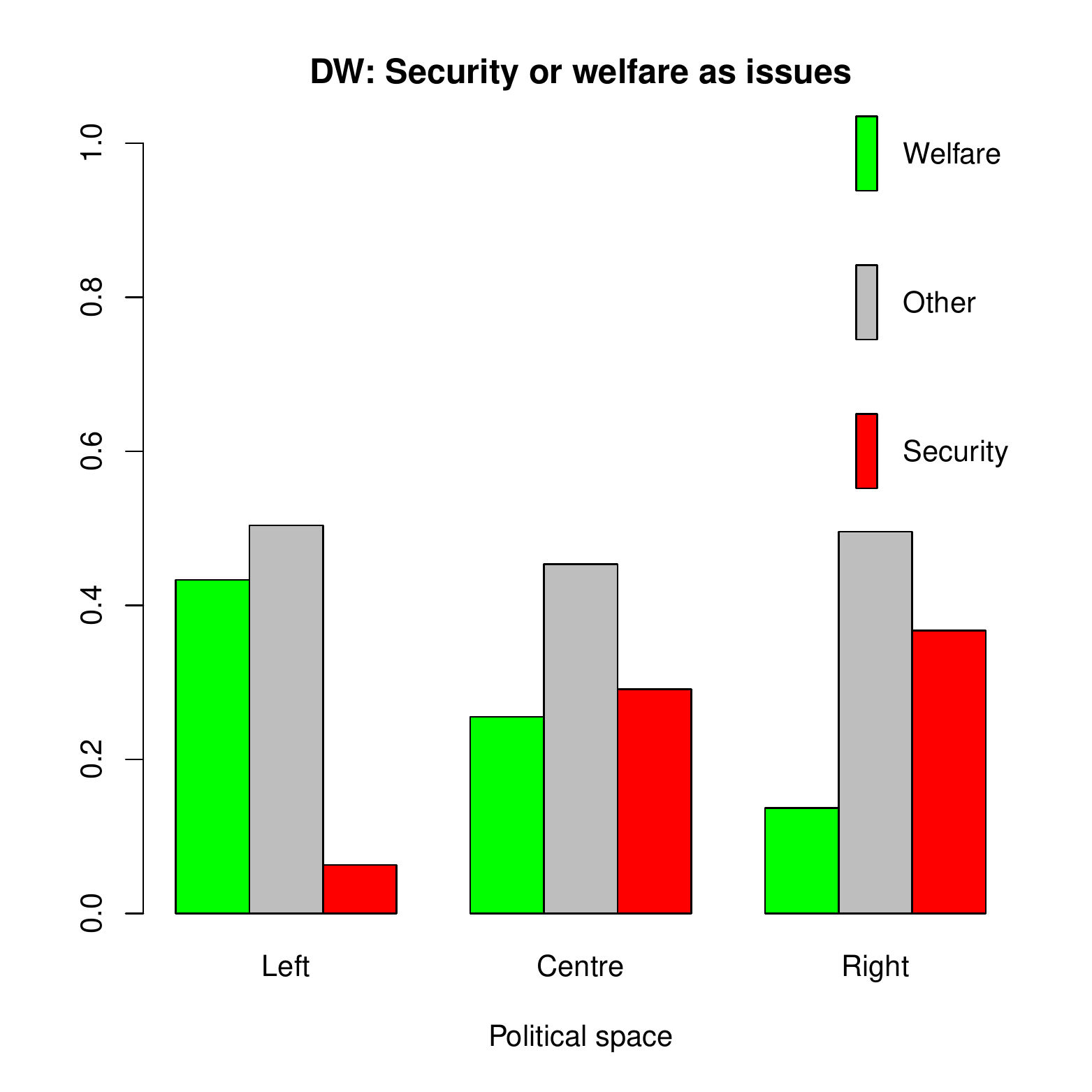}
         \caption{German two-ways table.}
         \label{f3:t3}
     \end{subfigure}
		\caption{Issues histograms and two-ways tables.}
\end{figure}

In Figures (\ref{f3:h1}), (\ref{f3:h2}), (\ref{f3:h3}), we reported the histogram with the size of the clusters detected by modularity community detection. In the case of Italy, see Figure (\ref{f3:h1}), it can be seen that the greatest cluster is composed by respondents whose concerns revolve around Economic issues, to the point that people worried about Security or Welfare appear as residual. Conversely, there are many respondents that were unclassified, as they are represented by arcs belonging to two different groups. So, we simplify the analysis considering respondents as they belong to just two groups: the economic group or the other, composed of the unclassified, the welfare, and the security. Next, we calculate all the two-ways table crossing the recoded concerns with social cleavages and we calculate their significance by the $p$-values.
We found that the most important cleavage that determines an economic concern is represented by the social class, whose $p$-value is 0.017. In Figure (\ref{f3:t1}) conditional frequencies are reported and it can be seen that the lower class is more worried by Economic issues than the middle class, even though Economic issues are at the core of the overall Italian concerns.

In Figures (\ref{f3:h2})  and (\ref{f3:h3}) histograms about the Spanish and the German clusters are reported too. In the case of Spain, the greatest cluster is represented by the Economic/Welfare concerns, still the Security cluster is not negligible, as it contains 7.6\% of the respondents. Therefore we continue the analysis leaving the two groups and the unclassified defined as above. The most significant cleavage describing concerns is the political position, for which the $p$-value is 0.064. In Figure (\ref{f3:t2}), conditional frequencies are reported: it can be seen that concerns about security increase from left to right wing voters, while concerns about economy or welfare decrease from left to right wing voters. In the case of Germany, the economy is the smallest cluster having only 1.6\% of respondents, therefore we aggregate this cluster to the unclassified to remain with security and welfare cluster only. The most significant cleavage describing concerns is political position, for which the $p$-value is $1.2(10)^{-9}$. In Figure (\ref{f3:t3}) conditional frequencies are reported and it can be seen that Security concern increases from left to right wing voters, while Welfare concern decreases from left to right wing voters, as was the case of Spain. 

In conclusion, the exercise shows how information of the items graph can be retrieved and used to determine what are the issues that are at the core of national public opinion's concern. We have seen that clusters can be interpreted as expressions of broad and general latent variables, that can be used to reduce data dimensionality. From the many possible pairs, or even the single alternatives only (that are 16), we have remained with two or three classes to which standard statistical analysis can be reliably applied. In conclusion, we think that the items graphs can be a convenient model to represent data coming from multiple issues questions, and a useful tool to complement or improve the statistical techniques used so far. 

\section{Conclusion and future research} 

In this contribution, we provided a network model to represent survey questions with multiple selections and we showed how to apply community detection to cluster the question items.  Community detection is only one of the many techniques that are used for network analysis, so, once that (hopefully) we have demonstrated the utility of the items graph, other network technique may be applied to it as well, such as centrality measures, core-periphery decomposition, and so on. One peculiar mention is deserved by the so-called overlapping community detection model, in which one node can belong to more than one community, see \cite{Xie2013}, in which the model is surveyed, and \cite{Benati2022} in which the mathematical model to calculate optimal communities are developed. In our case, this model is helpful to reduce the number of respondents that were classified as {\it other} in the last steps of the analysis for the plain fact that that their choices belong to two different groups. If communities/groups can overlap, then more arcs can belong to one of the groups and respondents could be better classified. A second possibility comes straight from the Eurobarometer application, as actually there are three questions about what are the most important issues. They are most important issues for the nation, personally, and for the European union, and so we have three items graphs. The three graph are connected by the fact that they contain the same item list and one respondent is represented by three arcs, one for each graph so what is obtained is a multilayer graph, to which specific techniques can be applied, see \cite{Mucha2010,Dickison2016}. Finally, there are surveys with questions in which respondents can elicit more than two items. For example in the same Eurobarometer we have analyzed, there is a question about what makes a sense of community between European citizens in which the items are: History, Religion, Values, Geography and many others, among which respondents can elicit up to {\it three} items. In this case, the representing answers by arcs is not sufficient as an arc can connect only two items. However, an arc can be readily extended to be an hyper-arc, that is, an arc connecting more than two nodes in the so-called hypergraphs, for which modularity optimization can be applied as well, see \cite{Kaminski2019, Kumar2020}. 

\section*{Acknowledgements}
We acknowledge the useful comments of our colleague Antonio Rodriguez-Chia during the development of this contribution. Moreover, the authors of this research have been supported by the financial aid of NetMeetData: Ayudas Fundacion BBVA a equipos de investigacion cientifica 2019.

\bibliographystyle{chicago}
    
\bibliography{references}

\end{document}